\documentclass[twocolumn,showpacs,preprintnumbers,amsmath,amssymb,nofootinbib]{revtex4}

\usepackage{graphicx}
\usepackage{dcolumn}
\usepackage{bm}
\usepackage{multirow}
\usepackage{graphicx}
\usepackage{subfigure}
\usepackage{color}
\usepackage{longtable}

\begin{document}


\title{Global Analysis of Nucleon Strange Form Factors at Low $Q^2$}

\author{Jianglai~Liu}
\email{jliu@caltech.edu}
\author{Robert~D.~McKeown}%
\affiliation{W.~K.~Kellogg Radiation Laboratory, California
Institute of Technology, Pasadena, CA 91125}
\author{Michael~J.~Ramsey-Musolf}
\altaffiliation{On leave from Department of Physics,
University of Wisconsin-Madison
Madison, WI 53706 USA}
\affiliation{W.~K.~Kellogg Radiation Laboratory, California
Institute of Technology, Pasadena, CA 91125}

%


\date{\today}

\begin{abstract}
We perform a global analysis of all recent experimental data from
elastic parity-violating electron scattering at low $Q^2$. The
values of the electric and magnetic strange form factors
of the nucleon are
determined at $Q^2 =  0.1$~GeV/$c^2$ to be $G^s_E = -0.008 \pm 0.016$ and
$G^s_M = 0.29 \pm 0.21$.
\end{abstract}

\maketitle
The existence of a ``sea'' of quarks and antiquarks in the nucleon
has been firmly established in deep-inelastic lepton scattering
experiments as well as in the production of dilepton pairs (the
Drell-Yan process). However, demonstrating the role of these $\bar q
q$ pairs in the static electromagnetic properties of the nucleon has
been a more elusive and difficult task.

As the lightest quark that contributes only to the $q{\bar q}$ sea,
the strange quark provides a unique window on the role of the sea in
the nucleon's electromagnetic structure.
As suggested by Kaplan and Manohar \cite{kaplan}, knowledge of
neutral current form factors, when combined with electromagnetic
form factors, provides access to the contribution of strange quarks
to these form factors. At low momentum transfers, the neutral
current form factors can be determined through parity-violating (PV)
electron scattering experiments \cite{bmck89,beck89}.

During the last decade, there has been dramatic progress in the
study of the strange quark-antiquark contributions to the nucleon
elastic electromagnetic form factors. A series of definitive
PV electron scattering experiments along with several
theoretical studies now provide a basis for extracting precision
information on these strange quark contributions. In this work we
report the results of a global analysis of all these experiments,
including
both the latest data obtained in experiments performed at the
Jefferson Laboratory and appropriate theoretical input on radiative
corrections, and obtain values for the strange electric and
magnetic form factors of the nucleon at a four-momentum transfer
$Q^2 =  0.1$~GeV/$c^2$.
We have also studied the sub-leading $Q^2$ dependence
of these two form factors, and find that so far the data do
not provide conclusive information.

\section{\label{sec_formalism} Strange Form Factors and Parity-violating
  Electron Scattering}
The nucleon vector strange form factors, $G_E^s$ and $G_M^s$, characterize
the contribution of the strange sea quarks to the nucleon electromagnetic
form factors, and thereby their contribution to the charge
and magnetization distributions in the nucleon.
With polarized electron facilities,
$G_E^s$ and $G_M^s$ can be accessed by measuring
the PV asymmetries in elastic $e$-$p$ scattering, 
quasielastic $e$-$d$ scattering, and elastic
$e$-$^4$He scattering~\cite{Mus94}.
In very general terms, the parity-violating
asymmetry $A_{PV}$ can be written as
\begin{equation}
  \label{eq_linear_comb}
  A_{PV} = A_{nvs} + \eta_E G_E^s + \eta_M G_M^s\,,
\end{equation}
where $A_{nvs}$ is the ``non-vector-strange'' asymmetry (independent
of $G_E^s$ and $G_M^s$), and $\eta_E$ and $\eta_M$ are functions
of kinematic quantities, nucleon electromagnetic form factors,
and nuclear models (for non-hydrogen targets).

For elastic e$-$p scattering, the full form of the asymmetry is~\cite{Mus94}
\begin{align}
\label{eq_Apv_full2}
\nonumber A_{PV}^{p} &= -\displaystyle\frac{G_FQ^2}{4\sqrt{2}\pi\alpha}
\frac{1}{\left[\epsilon(G_E^{p})^2 + \tau(G_M^{p})^2\right]}\\
\nonumber
&\times \{\displaystyle
(\epsilon(G_E^{p})^2 + \tau(G_M^{p})^2)(1-4\sin^2\theta_W)(1+R_V^p)\\
\nonumber
&\quad -(\epsilon G_E^p G_E^n + \tau G_M^p G_M^n)(1+R_V^n)\\
\nonumber
&\quad -(\epsilon G_E^p {\color{blue}G_E^s} + \tau G_M^p {\color{blue}G_M^s})(1+R_V^{(0)})\\
&\quad -\epsilon'(1-4\mathrm{sin}^2\theta_W)G_M^pG_A^e\}\,,
\end{align}
with
\begin{align}
  \nonumber
  &\tau = \displaystyle\frac{Q^2}{4M_p^2}\,, \quad
  \epsilon = \left(1+2(1+\tau)\mathrm{tan}^2\frac{\theta}{2}\right)^{-1}\,,\\
  \nonumber
  &\epsilon'= \displaystyle\sqrt{\tau(1+\tau)(1-\epsilon^2)}\,,
\end{align}
where $M_p$ is the mass of the proton and $\theta$ is the electron scattering
angle.
In Eqn.~\ref{eq_Apv_full2},
$G_F$ and $\alpha$ are the Fermi and fine structure constants,
respectively.
$Q^2$ is the four momentum transfer. $G_{E,M}^{(p,n)}$
are the proton and neutron electric and magnetic form factors, while
$G_A^e$ is proton axial form factor seen by an electron. In order to
extract contributions from $G_{E,M}^s$ to $A^p_{PV}$, one must include
the effects of Standard Model (SM) ${\cal O}(\alpha)$ electroweak
radiative corrections~\cite{KubisComment}.
It is often useful 
to characterize these corrections
in terms of ratios $R_{V,A}$ of the ${\cal O}(\alpha)$ hadronic vector ($V$)
and axial vector ($A$) weak neutral current amplitudes to the corresponding
tree-level amplitudes. The  $R_V^p$, $R_V^n$, $R_V^{(0)}$ give these ratios
for vector proton, neutron, and SU(3)-singlet amplitudes, respectively. In
principle, their values can be obtained using the SM predictions for the
effective electron-quark couplings $C_{1q}$ given in  \cite{PDG06}. However,
the quoted $C_{1q}$ do not include perturbative QCD contributions or
coherent strong interaction effects in the radiative corrections associated
with elastic scattering from a nucleon. A recent analysis of these effects
has been given in Ref.~\cite{Erler:2003yk} and up-dated in
Ref.~\cite{Erler:2004in}. The latter work also gives an improved
treatment of strong interaction contributions to the running of the
weak mixing angle in the $\overline{MS}$ renormalization scheme
from its value at the $Z$-pole,  ${\hat s}_Z^2\equiv\sin^2\hat\theta_W(M_Z)$,
to the quantity appropriate for precise, low-energy neutral current
experiments, $\sin^2\hat\theta_W(0)$. All of these effects are included
in the $R_V$
given in Table \ref{tab_ana_prms}. 
The theoretical uncertainties in
 $R_V^{n}$ and $R_V^{(0)}$ are less than one percent and have a negligible
impact on our analysis, so we do not quote these errors in
Table \ref{tab_ana_prms}. The theoretical error in
$(1-4\sin^2\theta_W)(1+R_V^p)$ is $\pm 0.0008$ \cite{Erler:2004in},
or slightly more than one percent. Since this error receives roughly
equal contributions from the uncertainty in ${\hat s}_Z^2$ and from
the ${\cal O}(\alpha)$ $Z\gamma$ box graph corrections, it is not
appropriate to quote an uncertainty in $R_V^p$ alone. The
uncertainties associated with other strong interaction effects
are sub-dominant. For the kinematic range of the PV experiments
analyzed here, the $R_V$  have a negligible impact on the
$Q^2$-dependence of $A_{PV}^p$ and are taken to be constant.
We use the conventional $\overline{MS}$ renormalization scheme. 
Therefore, $\sin^2\theta_W$ in Eqn.~\ref{eq_Apv_full2}
and hereafter shall take its value as ${\hat s}_Z^2$.

For reader's convenience, constant parameters (in some cases 
also the uncertainties) appearing in Eqn.~\ref{eq_Apv_full2} 
and later in Eqns.~\ref{eq_GAe} and \ref{eq:Gdipole} are 
summarized in Table~\ref{tab_ana_prms}.
\begin{table}[!ht]
  \renewcommand{\arraystretch}{1.2}
  \centering
  \begin{tabular}{ll} \\
    \begin{tabular}{cc} \hline
      Parameter & Value \\\hline
      $\alpha$  & 1./137.03599911 \\
      $\hat s^2_Z$ & 0.23122(15)\\
      $G_F$        & $1.16637\times10^{-5}$/GeV$^2$\\
      $M_p$         & 0.98272~GeV \\
      $\Lambda_A^2$ & 1.00(0.04)~(GeV/c)$^2$ \\\hline
    \end{tabular}
    & 
    \begin{tabular}{cc} \hline
      Parameter & Value\\\hline
      $g_A/g_V$    & $-$1.2695\\
      $3F-D$    & 0.58(0.12)\\
      $\Delta s$ & $-$0.07(0.06)\\
      $R_V^p$   & $-$0.0520 \\
      $R_V^n$   & $-$0.0123  \\
      $R_V^{(0)}$ & $-$0.0123 \\\hline
    \end{tabular}
  \end{tabular}
  \caption{Summary of parameters (with some uncertainties in parentheses)
    in Eqns.~\ref{eq_Apv_full2}, \ref{eq_GAe}, and \ref{eq:Gdipole}.
    The values of $\alpha$, $\hat s^2_Z$, $G_F$, $M_p$,
    $g_A/g_V$ are taken directly from \cite{PDG06}.
    $R_V^n$, and $R_V^{(0)}$ are converted from $C_{1q}$ parameters
    in \cite{PDG06}. $R_V^p$ is derived from the
    proton weak charge given in Ref.~\cite{Erler:2004in}. We adopt the 
    value and uncertainty of $\Lambda_A^2$ from \cite{BBA03}, 
    $3F-D$ from \cite{Fil01}, and $\Delta s$ from \cite{Ada97}.
  }
  \label{tab_ana_prms}
\end{table}

The effective axial form factor $G_A^e$ receives a number of contributions and may be written as
\begin{align}
  \label{eq_GAe}
  \nonumber
  &\displaystyle G_A^e(Q^2) = G_D(Q^2)\times\\
  &\left[\frac{g_A}{g_V}(1+R_A^{T=1})+\frac{3F-D}{2}R_A^{T=0}
    +\Delta s (1+R_A^{(0)})\right]\,,
\end{align}
where
\begin{equation}
  \label{eq:Gdipole}
  G_D(Q^2) = \frac{1}{(1+Q^2/\Lambda_A^2)^2}\,,
\end{equation}
parameterizes the $Q^2$-dependence with a dipole form and
 $\Lambda_A$ is the corresponding  axial dipole mass. The ratio $-\frac{g_A}{g_V}$ is the
isovector axial form factor of the nucleon
at zero momentum transfer, which is precisely measured in the neutron beta
decay. $F$ and $D$ are the octet baryon beta decay parameters, which
can be determined by combining data from neutron and hyperon beta
decays under the assumption of SU(3) flavor symmetry. $\Delta s$
is the strange quark contribution to nucleon spin. Assuming a gentle
evolution from the perturbative to the non-perturbative domain, this
quantity can be obtained from inclusive, polarized deep inelastic
lepton-nucleon scattering.

The ratios $R_A^{T=1}$, $R_A^{T=0}$, and $R_A^{(0)}$ characterize
the effects of electroweak radiative corrections to the isovector,
isoscalar, and SU(3) singlet hadronic axial vector amplitudes. Note
that while $R_A^{T=1}$ and $R_A^{(0)}$ give the ratios of the ${\cal
O}(\alpha)$ and tree-level hadronic axial vector neutral current
amplitudes in the isovector and SU(3) singlet channels,
respectively, $R_A^{T=0}$ does not have this interpretation since
the tree-level isoscalar hadronic axial vector amplitude vanishes in
the SM.

Conventionally, these  quantities are divided into two pieces: the
``one-quark'' and ``many-quark'' contributions. The one-quark
contributions correspond to renormalization of the effective vector
electron-axial vector quark couplings, $C_{2q}$, and their values
can be obtained from the SM predictions for these couplings given in
Ref.~\cite{PDG06}. The many-quark contributions include the
so-called \lq\lq anapole" effects as well as coherent strong
interaction contributions to the radiative corrections. In contrast
to the situation with the vector corrections, $R_V$, the {\em
relative} importance of many-quark effects in the $R_A$ can be quite
pronounced. The small vector coupling of the electron to the
$Z$-boson, $g_V^e=-1+4\sin^2\hat\theta_W\sim-0.075$, leads to a
suppression of the tree-level hadronic axial vector amplitude.
However, $g_V^e$  is absent from a variety of both one- and
many-quark radiative corrections. Thus, one would expect the
magnitudes of the $R_A$ to be of order several percent, rather than
the generic $\alpha/\pi\sim 0.3\%$ scale normally associated with
electroweak radiative corrections. As a result, the impact of
otherwise negligible strong interaction effects in the many-quark
corrections, such as the anapole contributions, can be
amplified~\cite{MRMOneQuark}.

An appropriate framework for treating the many-quark effects
associated with physics at the hadronic scale is chiral perturbation
theory. A comprehensive analysis of the anapole contributions to
$R_A^{T=1}$ and $R_A^{T=0}$  has been carried out to chiral order $p^3$
in Ref.~\cite{Zhu00}. This analysis included both one-loop
contributions associated with the octet of pseudoscalar mesons
as well as the full set of low-energy constants (LECs) that arise
at this chiral order. A generous theoretical range was assigned to
the LECs, leading to a quoted theoretical uncertainty in the total
$R_A$ that is larger than the  (logarithmically enhanced) one-quark
corrections. The theoretical SM uncertainty is likely to be smaller.

The corresponding results, updated for the present value of the
weak mixing angle, are given in Table \ref{tab_anapole_R}. The
resulting prediction for $G_A^e$ is consistent with both the results
of the SAMPLE deuterium
measurement~\cite{Ito04,Bei05}, which is particularly sensitive to
the dominant isovector axial component, as well as other theoretical
models for the anapole contributions~\cite{Riska:2000qw,Maekawa:2000qz}.
No evaluation of the ``many-quark'' contribution of $R_A^{(0)}$
has been made in the literature. We assume it is zero and assign
the size of  ``one-quark'' value for $R_A^{(0)}$ as its uncertainty.
\begin{table}[!ht]
  \centering
  \renewcommand{\arraystretch}{1.2}
  \begin{tabular}{cccc} \hline
    & $R_A^{T=1}$ & $R_A^{T=0}$ & $R_A^{(0)}$ \\\hline
    one-quark & $-0.172$ & $-0.253$ & $-0.551$ \\
    many-quark & $-0.086(0.34)$ & 0.014(0.19) & N/A \\
    total      & $-0.258(0.34)$ & $-0.239(0.20)$ & $-0.55(0.55)$\\\hline
  \end{tabular}
  \label{tab_anapole_R}
  \caption{The ``one-quark''~\cite{PDG06} and ``many-quark''~\cite{Zhu00}
    corrections to the axial charges, both in $\overline{MS}$,
    as well as the combined corrections.}
\end{table}

The PV asymmetry for the neutron can be obtained by exchanging the
``$p$'' and ``$n$'' indices on nucleon form factors in
Eqn.~\ref{eq_Apv_full2}, and flipping the sign of the first
isovector term in the expression for $G_A^e$ in Eqn.~\ref{eq_GAe}.
To first order, the PV asymmetry from a deuterium target is a
cross-section weighted average of the proton and neutron
asymmetries, which leads to an enhancement of the contribution of
$G_A^e$ and a suppression to the relative contribution due to
$G_E^s$ and $G_M^s$. Obviously a nuclear correction needs to be
applied in the analysis. In this note, for the SAMPLE deuterium
measurement, we shall adopt the asymmetry expression given 
in~\cite{Ito04}.

The $^4$He nucleus is spin zero, parity even and isoscalar. The PV asymmetry
takes a much simpler form~\cite{Mus94}:
\begin{align}
\label{eq_He}
\nonumber
A_{PV}^{He} &= \displaystyle{\frac{G_FQ^2}{4\pi\sqrt{2}\alpha}}\\
&\times\left(4\sin^2\theta_W(1+R_V^{T=0})
+ \frac{2(1+R_V^{(0)})G_E^s}{G_E^p+G_E^n}\right)\,,
\end{align}
where the isoscalar $R_V$ factor is related
to $R_V^p$ and $R_V^n$ as
\begin{equation}
  \displaystyle R_V^{T=0}
  = \frac{R_V^n-(1-4\sin^2\theta_W)R_V^p}{4\sin^2\theta_W}\,.
\end{equation}

\section{\label{sec_experiment} Experimental Data}
In this section, the world data of PV elastic scattering within
a $Q^2$ range from 0.07 to 0.5~(GeV/c)$^2$
will be summarized. These include
SAMPLE-H~\cite{Spa00,Bei05}, SAMPLE-D~\cite{Ito04,Bei05}, HAPPEx-H-99~\cite{Ani04},
HAPPEx-H-a~\cite{Ani05}, HAPPEx-He-a~\cite{Ani05b}, HAPPEx-H-b and
HAPPEx-He-b~\cite{Ach07}, PVA4-H-a~\cite{Maa04}, PVA4-H-b~\cite{Maa05},
and the first 14 $Q^2$ bins in $G^0$ forward angle~\cite{Arm05}.
The kinematics, targets, and the measured
asymmetries in these experiments are summarized in
Table~\ref{tab_world_data2}. In column $A_{phys}$, the first and
second uncertainties for the $G^0$ data are the uncorrelated and
correlated experimental uncertainties, respectively.
The values of $\eta_E$ and $\eta_M$ are also listed in the table.
In calculating them, we have adopted a recent parametrization
of the nucleon electromagnetic form factors
from Ref.~\cite{Kel04}. For the SAMPLE deuterium measurement, the
$\eta_M$ is taken from Ref.~\cite{Ito04}, whereas its $\eta_E$
is taken to be $1.79$ according to the static approximation.
\begin{table*}[!ht]
  \renewcommand{\arraystretch}{1.2}
  \addtolength{\tabcolsep}{1 mm}
  \begin{tabular}{cccccccccc} \hline
    \multirow{2}{*}{Experiment} &\multirow{2}{*}{Target}& $Q^2$ & $\theta_{lab}$
    & $A_{phys}$ & {$\eta_E$} & {$\eta_M$ }
    & \multirow{2}{*}{Reference}\\
    & &  (GeV/c)$^2$ &  (deg) & (ppm) & (ppm) & (ppm) &\\\hline
    SAMPLE-H & H$_2$ & 0.1 & 144.4 & $-5.61\pm1.11$
    & 2.07 & 3.47 &\cite{Spa00,Bei05}\\
    SAMPLE-D & D$_2$ & 0.091 & 140.8 & $-7.77\pm1.03$
    & 1.79 & 0.77 & \cite{Ito04,Bei05}\\\hline
    HAPPEx-H-99 & H$_2$ & 0.477 & 12.3 & $-15.05\pm1.13$
    & 56.89 & 22.62 & \cite{Ani04}\\
    HAPPEx-H-a & H$_2$ & 0.099 & 6.0 & $-1.14\pm0.25$
    & 9.55 & 0.76 & \cite{Ani05}\\
    HAPPEx-H-b & H$_2$ & 0.109 & 6.0 & $-1.58\pm0.13$
    & 10.59 & 0.93 & \cite{Ach07}\\
    HAPPEx-He-a & $^4$He & 0.091 & 5.7 & $6.72\pm0.87$
    & 20.19 & 0 & \cite{Ani05b}\\
    HAPPEx-He-b & $^4$He & 0.077 & 5.8 & $6.40\pm0.26$
    & 16.56 & 0 & \cite{Ach07}\\\hline
    PVA4-H-b & H$_2$ & 0.108 & 35.52 & $-1.36\pm0.32$
    & 10.08 & 1.05 & \cite{Maa05}\\
    PVA4-H-a & H$_2$ & 0.23 & 35.45 & $-5.44\pm0.60$
    & 22.56 & 5.07 & \cite{Maa04}\\ \hline
    $G^0$ & H$_2$ & 0.122 & 6.68 & $-1.51\pm0.49\pm0.18$
    & 11.96 & 1.17 & \cite{Arm05}\\
    $G^0$ & H$_2$ & 0.128 & 6.84 & $-0.97\pm0.46\pm0.17$
    & 12.60 & 1.30 & \cite{Arm05}\\
    $G^0$ & H$_2$ & 0.136 & 7.06 & $-1.30\pm0.45\pm0.17$
    & 13.46 & 1.48 & \cite{Arm05}\\
    $G^0$ & H$_2$ & 0.144 & 7.27 & $-2.71\pm0.47\pm0.18$
    & 14.32 & 1.66 & \cite{Arm05}\\
    $G^0$ & H$_2$ & 0.153 & 7.50 & $-2.22\pm0.51\pm0.21$
    & 15.31 & 1.89 & \cite{Arm05}\\
    $G^0$ & H$_2$ & 0.164 & 7.77 & $-2.88\pm0.54\pm0.23$
    & 16.53 & 2.19 & \cite{Arm05}\\
    $G^0$ & H$_2$ & 0.177 & 8.09 & $-3.95\pm0.50\pm0.20$
    & 17.99 & 2.58 & \cite{Arm05}\\
    $G^0$ & H$_2$ & 0.192 & 8.43 & $-3.85\pm0.53\pm0.19$
    & 19.69 & 3.07 & \cite{Arm05}\\
    $G^0$ & H$_2$ & 0.210 & 8.84 & $-4.68\pm0.54\pm0.21$
    & 21.77 & 3.71 & \cite{Arm05}\\
    $G^0$ & H$_2$ & 0.232 & 9.26 & $-5.27\pm0.59\pm0.23$
    & 24.37 & 4.60 & \cite{Arm05}\\
    $G^0$ & H$_2$ & 0.262 & 9.92 & $-5.26\pm0.53\pm0.17$
    & 28.00 & 5.99 & \cite{Arm05}\\
    $G^0$ & H$_2$ & 0.299 & 10.63 & $-7.72\pm0.80\pm0.35$
    & 32.60 & 7.99 & \cite{Arm05}\\
    $G^0$ & H$_2$ & 0.344 & 11.46 & $-8.40\pm1.09\pm0.52$
    & 38.40 & 10.89 & \cite{Arm05}\\
    $G^0$ & H$_2$ & 0.410 & 12.59 & $-10.25\pm1.11\pm0.55$
    & 47.28 & 16.10 & \cite{Arm05}\\
    \hline
  \end{tabular}
  \caption{A summary of the world data on PV elastic electron
    scattering within the range of
    0.07~(GeV/c)$^2<Q^2<$0.5~(GeV/c)$^2$, including the
    average kinematics, targets, published asymmetries $A_{phys}$,
    as well as coefficients
    $\eta_E$ and $\eta_M$. $A_{phys}$, $\eta_E$ and $\eta_M$ are
    in units of parts-per-million (ppm). The central kinematics for the
    two PVA4 measurements are obtained from \cite{AM05p}.
    For $A_{phys}$,
    the first and second uncertainties for the $G^0$ data are the
    uncorrelated and correlated experimental uncertainties, respectively.
  }
  \label{tab_world_data2}
\end{table*}
\section{\label{sec_analysis} Global Analysis}
\subsection{$A_{nvs}$ and Theoretical Uncertainties}
We shall now present a combined analysis of the world data aiming
to extract $G_E^s$ and $G_M^s$. A global fit, generally speaking, 
is obtained by simultaneously solving a set of equations
\begin{equation}
\label{eq_comb_fit_general}
  m_i\pm\sigma(m_i) = t_i(a_1,a_2,\cdots)\pm\sigma(t_i)\,,
\end{equation}
where $m_i$ and $t_i(a_1,a_2,\cdots)$, respectively, are the
measured and theoretical values for experiment $i$. In this
expression, $\sigma(m_i)$ and $\sigma(t_i)$ are their uncertainties,
and $a_1,a_2,\cdots$ are the free parameters one seeks to determine.
In our case,
\begin{equation}
\label{eq_comb_fit_strange}
  m_i = A_{phys}^i\,, t_i = A_{nvs} + \eta_E G_E^s + \eta_M G_M^s\,,
\end{equation}
with $G_E^s$ and $G_M^s$ being the free parameters.
In the previous section, we have discussed the value and uncertainty
of $A_{phys}^i$, as well as $\eta_E$ and $\eta_M$ (Table~\ref{tab_world_data2}).
For each measurement, the values of $A_{nvs}$ can be also computed
straightforwardly using the formalism in Sec.~\ref{sec_formalism}. They
are listed in Table~\ref{tab_err_Anvs}. We again have
used the parametrization of the
nucleon electromagnetic form factors from Ref.~\cite{Kel04}. As mentioned, 
the $A_{nvs}$ for the SAMPLE deuterium measurement is calculated 
based on the asymmetry expression in \cite{Ito04} with the theoretical
value of $G_A^e$.

The treatment of the theoretical uncertainties $\sigma(t_i)$ is
more subtle. $\sigma(t_i)$ receives dominant contributions from the
following sources: the nucleon axial form factor ($G_A^e$), nucleon
electromagnetic form factors ($G_{E,M}^{(p,n)}$), and nuclear
corrections. Theoretical uncertainties from a given source are
correlated among different experiments. The uncertainty in $G_A^e$
can be calculated based on the uncertainties in
Tables~\ref{tab_anapole_R} and \ref{tab_ana_prms}, and is dominated
by the uncertainty of the ``many-quark'' electroweak radiative
corrections on the $R_A$ factors in Table~\ref{tab_anapole_R}. For
the nucleon electromagnetic form factors, based on the spread of the
world data (see, e.g. Ref.~\cite{Kel04}), we estimate their relative
uncertainties as
\begin{align}
  \nonumber
  &\frac{\sigma(G_E^p)}{G_E^p}=2.5\%\,,\frac{\sigma(G_M^p)}{G_M^p}=1.5\%\,,\\
  \nonumber
  &\frac{\sigma(G_E^n)}{G_E^n}=15\%\,\,\rm{and}\,\frac{\sigma(G_M^n)}{G_M^n}=1\%\,,
\end{align}
respectively. This is consistent with the uncertainty assignment in
Ref.~\cite{Ani05}, except that the uncertainty of $G_E^n$ in
\cite{Ani05} was assigned more conservatively to be 30\%. Also note
that we have made a simplifying assumption that these form factor
uncertainties are ``scaling'' in nature, independent of the $Q^2$.
For an analysis with relatively small range of $Q^2$, this is
reasonable. Nuclear corrections are only relevant for non-hydrogen
targets. For the  two $^4$He measurements, according to
Ref.~\cite{Ach07}, 3\% is assigned as the fractional theoretical
uncertainty of $A_{nvs}$. Nuclear corrections for the SAMPLE
deuterium measurement have little impact on the final fit, and are
therefore neglected. The theoretical uncertainties for the $t_i$
in Eqn.~\ref{eq_comb_fit_strange} due to the different sources are
summarized in the last six columns in Table~\ref{tab_err_Anvs}. To
be precise, the content of each column gives the change in $A_{nvs}
+ \eta_E G_E^s + \eta_M G_M^s$ when the source magnitude ($|G_A^e|$,
$|G_{E,M}^{(p,n)}|$, or nuclear correction to $^4$He data) is
increased  by one standard deviation. Notice that the nucleon
electromagnetic form factors also affect the value of $\eta_E$ and
$\eta_M$, therefore generate ``pull terms'' linear in $G_E^s$ and
$G_M^s$ in Table~\ref{tab_err_Anvs}. Such pull terms are neglected
for the SAMPLE deuterium measurement.
\begin{table*}[!ht]
  \renewcommand{\arraystretch}{1.2}
  \addtolength{\tabcolsep}{0.4 mm}
  \centering
{\small
  \begin{tabular}{ccc|c|c|c|c|c|c} \hline
    \multirow{2}{*}{Exp}&$Q^2$ & $A_{nvs}$
    & \multicolumn{6}{c}{Theoretical uncertainty $\sigma(t_i)$ (ppm)}\\\cline{4-9}
    &  (GeV/c)$^2$ & ppm & $G_A^e$ & $G_E^p$ & $G_M^p$ & $G_E^n$ & $G_M^n$ & N.C. \\\hline
    SAMPLE-H & 0.1  &$-6.85$ & $0.57$ & $0.06+0.03G_E^s-0.03G_M^s$   & $0.06-0.05G_E^s-0.03G_M^s$ & $0.01$  & $-0.05$ &0\\
    SAMPLE-D & 0.091 & $-$8.37  & 0.60 & 0.05 & 0.02 & 0.02 & $-$0.03 & 0 \\
    HAPPEx-H-99 & 0.477 &$-15.96$ & $0.23$ & $0.38+0.07G_E^s-0.52G_M^s$   & $-0.03-0.90G_E^s-0.02G_M^s$ & $0.49$  & $-0.16$ &0\\
    HAPPEx-H-a & 0.099 &$-1.42$ & $0.01$ & $0.04-0.15G_E^s-0.03G_M^s$   & $-0.01-0.05G_E^s+0.01G_M^s$ & $0.05$  & $-0.01$ &0\\
    HAPPEx-H-b & 0.109 &$-1.64$ & $0.02$ & $0.05-0.16G_E^s-0.04G_M^s$   & $-0.01-0.06G_E^s+0.01G_M^s$ & $0.06$  & $-0.01$ &0\\
    HAPPEx-He-a & 0.091 & 7.50 & 0 & $-0.47G_E^s$& 0& $-0.12G_E^s$& 0& 0.23 \\
    HAPPEx-He-b & 0.077 & 6.35 & 0 & $-0.39G_E^s$& 0& $-0.08G_E^s$& 0& 0.19 \\
    PVA4-H-b & 0.108 &$-2.02$ & $0.09$ & $0.06-0.14G_E^s-0.04G_M^s$   & $-0.02-0.07G_E^s+0.01G_M^s$ & $0.05$  & $-0.01$ &0\\
    PVA4-H-a & 0.23 &$-6.26$ & $0.26$ & $0.17-0.13G_E^s-0.15G_M^s$   & $-0.03-0.26G_E^s+0.02G_M^s$ & $0.18$  & $-0.05$ &0\\
    $G^0$  & 0.122  &$-1.94$ & $0.02$ & $0.06-0.17G_E^s-0.04G_M^s$   & $-0.02-0.08G_E^s+0.01G_M^s$ & $0.07$  & $-0.02$  &0\\
    $G^0$  & 0.128  &$-2.09$ & $0.02$ & $0.06-0.17G_E^s-0.05G_M^s$   & $-0.02-0.09G_E^s+0.01G_M^s$ & $0.07$  & $-0.02$  &0\\
    $G^0$  & 0.136  &$-2.29$ & $0.02$ & $0.07-0.18G_E^s-0.05G_M^s$   & $-0.02-0.10G_E^s+0.01G_M^s$ & $0.08$  & $-0.02$  &0\\
    $G^0$  & 0.144  &$-2.50$ & $0.03$ & $0.07-0.18G_E^s-0.06G_M^s$   & $-0.02-0.11G_E^s+0.01G_M^s$ & $0.09$  & $-0.02$  &0\\
    $G^0$  & 0.153  &$-2.74$ & $0.03$ & $0.08-0.19G_E^s-0.07G_M^s$   & $-0.02-0.12G_E^s+0.01G_M^s$ & $0.10$  & $-0.02$  &0\\
    $G^0$  & 0.164  &$-3.05$ & $0.04$ & $0.09-0.19G_E^s-0.08G_M^s$   & $-0.02-0.13G_E^s+0.01G_M^s$ & $0.11$  & $-0.03$  &0\\
    $G^0$  & 0.177  &$-3.43$ & $0.04$ & $0.10-0.19G_E^s-0.09G_M^s$   & $-0.03-0.16G_E^s+0.02G_M^s$ & $0.13$  & $-0.03$  &0\\
    $G^0$  & 0.192  &$-3.88$ & $0.05$ & $0.11-0.19G_E^s-0.10G_M^s$   & $-0.03-0.18G_E^s+0.02G_M^s$ & $0.14$  & $-0.03$  &0\\
    $G^0$  & 0.210  &$-4.46$ & $0.06$ & $0.13-0.19G_E^s-0.12G_M^s$   & $-0.03-0.21G_E^s+0.02G_M^s$ & $0.16$  & $-0.04$  &0\\
    $G^0$  & 0.232  &$-5.21$ & $0.07$ & $0.15-0.19G_E^s-0.15G_M^s$   & $-0.04-0.25G_E^s+0.02G_M^s$ & $0.19$  & $-0.05$  &0\\
    $G^0$  & 0.262  &$-6.30$ & $0.08$ & $0.18-0.18G_E^s-0.18G_M^s$   & $-0.04-0.32G_E^s+0.02G_M^s$ & $0.23$  & $-0.06$  &0\\
    $G^0$  & 0.299  &$-7.75$ & $0.11$ & $0.21-0.15G_E^s-0.23G_M^s$   & $-0.04-0.40G_E^s+0.02G_M^s$ & $0.27$  & $-0.07$  &0\\
    $G^0$  & 0.344  &$-9.65$ & $0.14$ & $0.26-0.11G_E^s-0.30G_M^s$   & $-0.04-0.51G_E^s+0.02G_M^s$ & $0.33$  & $-0.09$  &0\\
    $G^0$  & 0.410  &$-12.70$ & $0.19$ & $0.32-0.03G_E^s-0.40G_M^s$   & $-0.04-0.69G_E^s+0.00G_M^s$ & $0.41$  & $-0.12$ &0\\
    \hline
  \end{tabular}
}
\caption{A summary of $A_{nvs}$ and the theoretical uncertainties
  of individual measurements. The uncertainties are grouped by
  six different sources: $G_A^e$, $G_E^p$, $G_M^p$, $G_E^n$, $G_M^n$
  and the nuclear corrections (N.C.) to $^4$He data.
  See text for details.
}
\label{tab_err_Anvs}
\end{table*}
\subsection{Combined Analysis at $Q^2$=0.1~(GeV/c)$^2$}
\label{sec_comb_0.1} 
\begin{figure}[!ht]
  \centering
  \includegraphics[width=0.48\textwidth]{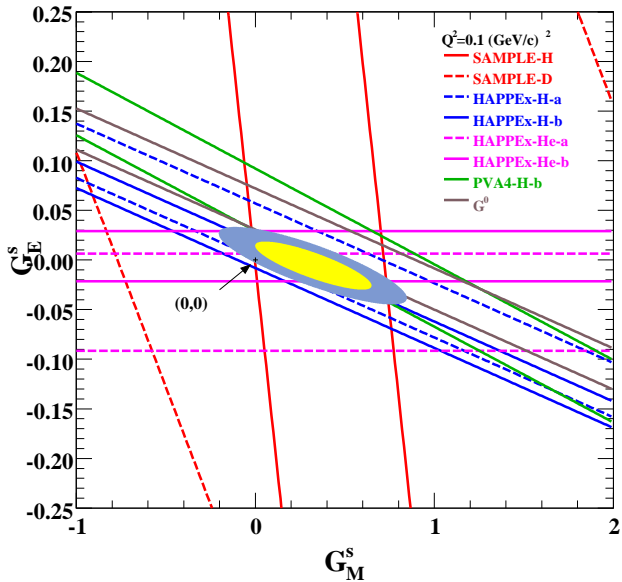}
  \caption{
    (Color online) The world data constraints on $(G_E^s,G_M^s)$
    at $Q^2=0.1$~(GeV/c)$^2$. The form factors of Kelly
    are used. Different bands in the plot represent:
    SAMPLE-H~\cite{Spa00} (solid red), SAMPLE-D~\cite{Ito04} (dashed red),
    HAPPEx-H-a~\cite{Ani05} (dashed blue),
    HAPPEx-H-b~\cite{Ach07} (solid blue),
    HAPPEx-He-a~\cite{Ani05b} (dashed pink),
    HAPPEx-He-b~\cite{Ach07} (solid pink),
    PVA4-H-b~\cite{Maa05} (solid green),
    and the lowest three $Q^2$ bins
    in $G^0$ forward angle~\cite{Arm05} (solid brown).
    The yellow and gray blue (dark) ellipses represents
    68.27\% ($\Delta \chi^2=2.3$) and 95\% ($\Delta \chi^2=5.99$)
    confidence contours around the point of maximum likelihood
    at ($G_E^s = -0.006, G_M^s = 0.33$). The black cross represents
    $G_E^s=G_M^s=0$.}
  \label{fig_bands}
\end{figure}

As seen from Table~\ref{tab_world_data2}, a
wealth of data exist with  $Q^2$ in the vicinity of 0.1~(GeV/c)$^2$,
including SAMPLE-H, SAMPLE-D, HAPPEx-H-a, HAPPEx-H-b, HAPPEx-He-a,
HAPPEx-He-b, PVA4-H-b, and low $Q^2$ data from $G^0$. It is natural
to first make a combined analysis at $Q^2 = 0.1$. To interpolate all
data to a common $Q^2$, we assume $G_E^s\propto
Q^2$ and $G_M^s$ is a constant~\cite{GesZero}.
That is, we replace $\eta_E
G_E^s(Q^2) + \eta_M G_M^s(Q^2)$ by ${\tilde\eta}_E G_E^s(0.1) +
\eta_M G_M^s(0.1)$, where $\tilde{\eta_E} =
\eta_E\frac{Q^2}{0.1~\rm{(GeV/c)^2}}$. To simplify our notation, we
shall use $G_E^s$ and $G_M^s$ hereafter to denote their values at
$Q^2$=0.1~(GeV/c)$^2$. In the $(G_E^s,G_M^s)$ space, each
measurement $i$ provides a linear constraint as
\begin{equation}
\label{eq_lincomb}
{\tilde\eta}_E^i G_E^s + \eta_M^i G_M^s
= A_{phys}^i-A_{nvs}^i\,.
\end{equation}
In Fig.~\ref{fig_bands}, each constraint is shown as a
linear band in the $(G_E^s,G_M^s)$ plane, where $\sigma(A_{phys}^i)$
(see Table~\ref{tab_world_data2}) and the theoretical uncertainty
$\sigma(t_i)$ (see Table~\ref{tab_err_Anvs}) have been combined in
quadrature into an overall uncertainty. Somewhat arbitrarily, we
include the 3 lowest $Q^2$ bins from $G^0$ data in this part of the
analysis. For visual clarity, they are combined into a single
constraint as
\begin{equation}
A_{phys}^i-A_{nvs}^i = 0.84 \pm 0.34 = 16.38 G_E^s
+ 1.32 G_M^s\,,
\end{equation}
which is shown as the solid brown band in Fig.~\ref{fig_bands}. From
the figure, one sees that the agreement among different measurements
is generally good. The $G^0$ and PVA4 appear to be offset from the
HAPPEx-H-b measurement, but they nevertheless agree within
2$\sigma$. As explained in Sec.~\ref{sec_formalism}, the SAMPLE
deuterium measurement (dashed red band) has much less sensitivity
to $G_E^s$ and $G_M^s$.

The 10 different measurements (3 from $G^0$, and the other 7 from separate
experiments) above provide redundancy in the joint
determination of $(G_E^s,G_M^s)$.
To solve for $(G_E^s,G_M^s)$ and determine the confidence contours,
we follow the standard least square procedure (see, e.g. Ref~\cite{Stump01}).
Specifically, we rearrange Eqn.~\ref{eq_lincomb} into the form
of Eqn.~\ref{eq_comb_fit_general} as
\begin{equation}
  \label{eq_least_square_recipe}
  m_i = t_i(G_E^s,G_M^s) \pm \sigma_i + \displaystyle{\sum_{k=0}^{6}}\pm\beta_{i,k}\,.
\end{equation}
where $m_i$ and $t_i$ are given in Eqn.~\ref{eq_comb_fit_strange},
and $\sigma_i$ is the uncorrelated experimental uncertainty.
$\beta_{i,k}$ denotes the correlated uncertainty for measurement $i$
with ``source index'' $k$. In our case, $\beta_{i,0}$ equals the
correlated experimental uncertainty for the $G^0$ data and 0 for
other experiments, and $\beta_{i,k=1,2,3,4,5,6}$ are the correlated
theoretical uncertainties for each measurement $i$ due to different
sources (Table~\ref{tab_err_Anvs}). Then for each given pair of
$(G_E^s,G_M^s)$, the $\chi^2$ is calculated as
\begin{equation}
\label{eq_chi2_corre}
  \chi^2 = \displaystyle \sum_{i} \sum_{j}
  (m_i-t_i)(V^{-1})_{ij} (m_j-t_j)\,,
\end{equation}
where $i$ and $j$ are indices of the measurements, and $V$ is the
variance matrix with $V_{ij} = \sigma_i^2\delta_{ij} +
\displaystyle\sum_{k=0}^{6}\beta_{i,k}\beta_{j,k}$. It has been
shown in Ref.~\cite{Stump01} that the $\chi^2$ constructed this way
satisfies the standard $\chi^2$ distribution, and the solution (best
fit) can be found by minimizing this $\chi^2$. Applying this
technique to the 10 measurements in Fig.~\ref{fig_bands}, we obtain
\begin{equation}
\label{eq_bestfit_band}
G_E^s = -0.006 \pm 0.016 \,, \quad G_M^s = 0.33 \pm 0.21
\end{equation}
with a correlation coefficient of $-0.83$ between the two, and a
minimum $\chi^2_{min} = 9.90$ with 8 degrees of freedom. Note that
the uncertainties above are 1$\sigma$ ($\Delta\chi^2=1$)
``marginalized'' uncertainties corresponding to the projections of
the error ellipse onto the two axes. That is, for a given value of
$G_M^s$, the $\chi^2$ is minimized by varying $G_E^s$ and vice
versa. The range defined by this uncertainty for a given parameter
corresponds precisely to 68.27\% of confidence interval of that
parameter~\cite{Bev92}. On the other hand, for the two parameters
($G_E^s$ and $G_M^s$) that are jointly determined, the 68.27\%
confidence region is instead defined by $\Delta\chi^2=2.3$
contour~\cite{PDG06}.
To demonstrate the precision of the fit, we plot the
68.27\% ($\Delta\chi^2=2.30$) and 95\% ($\Delta\chi^2=5.99$)
joint confidence levels in Fig.~\ref{fig_bands}
as the yellow (light) and gray (dark) ellipses, respectively.
$(G_E^s,G_M^s)=(0,0)$ yields a $\Delta\chi^2$ of 4.81
($\int_{4.8}^{\infty}P_{\chi^2}d\chi^2 = 9.0\%$ where $P_{\chi^2}$ is 
the $\chi^2$ probability distribution function~\cite{Bev92}), 
which is depicted as the black cross in the figure.

\subsection{Inclusion of higher $Q^2$ data from $G^0$, PVA4, and HAPPEx}
The analysis presented so far has focused on the vicinity of
$Q^2$=0.1~(GeV/c)$^2$. It is desirable to extend the analysis by
including data at higher $Q^2$ from $G^0$, PVA4, and HAPPEx. As a
first attempt, the previous assumption that $G_E^s\propto Q^2$ and
$G_M^s$ is a constant was adopted. We shall refer to such a fit as
the first order fit. Using the same $\chi^2$ construction as in
Eqns.~\ref{eq_least_square_recipe} and \ref{eq_chi2_corre}, we
started by fitting the data in Table~\ref{tab_world_data2} with
$Q^2<$0.11~(GeV/c)$^2$ (up to PVA4-H-b), and then systematically
included more data in the fit with increasing $Q^2$. In
Table~\ref{tab_fit_class1}, the resulting $(G_E^s,G_M^s)$ and the
fit quality are summarized.
\begingroup
\squeezetable
\begin{table}[!ht]
  \renewcommand{\arraystretch}{1.4}
  \centering
  \begin{tabular}{cccccc}
    \hline
    $Q^2$ & \multirow{2}{*}{{\small Experiment}}
    & \multirow{2}{*}{$G_E^s$} & \multirow{2}{*}{$G_M^s$}
    & \multirow{2}{*}{$\chi_{min}^2/\nu$}
    & Prob\\
    (GeV/c)$^2$ & & & & & (\%) \\\hline
    \multicolumn{2}{c}{$<$0.11~(GeV/c)$^2$}
    & $-$0.008(0.016) & 0.26(0.22) & 4.7/5 & 45.5\\\hline
    0.122 & $G^0$ & $-$0.007(0.016) & 0.27(0.22) & 4.9/6 & 55.4\\
    0.128 & $G^0$ & $-$0.006(0.016) & 0.30(0.22) & 8.4/7 & 29.7\\
    0.136 & $G^0$ & $-$0.006(0.016) & 0.33(0.21) & 9.9/8 & 27.2\\
    0.144 & $G^0$ & $-$0.007(0.016) & 0.30(0.21) & 13.0/9 & 16.1\\
    0.153 & $G^0$ & $-$0.007(0.016) & 0.30(0.21) & 13.0/10 & 22.1\\
    0.164 & $G^0$ & $-$0.008(0.016) & 0.29(0.21) & 13.8/11 & 24.4\\
    0.177 & $G^0$ & $-$0.010(0.016) & 0.24(0.21) & 19.5/12 & 7.7\\
    0.192 & $G^0$ & $-$0.010(0.016) & 0.22(0.21) & 20.3/13 & 8.9\\
    0.210 & $G^0$ & $-$0.012(0.016) & 0.19(0.21) & 22.1/14 & 7.6\\
    0.230 & PVA4  & $-$0.013(0.016) & 0.22(0.20) & 22.3/15 & 10.1\\
    0.232 & $G^0$ & $-$0.014(0.016) & 0.21(0.20) & 23.1/16 & 11.0\\
    0.262 & $G^0$ & $-$0.012(0.016) & 0.24(0.19) & 24.7/17 & 10.2\\
    0.299 & $G^0$ & $-$0.014(0.016) & 0.23(0.19) & 26.0/18 & 10.0\\
    0.344 & $G^0$ & $-$0.014(0.016) & 0.24(0.19) & 26.1/19 & 12.7\\
    0.410 & $G^0$ & $-$0.013(0.015) & 0.27(0.19) & 27.1/20 & 13.1\\
    0.477 & HAPPEx-H-99 & $-$0.015(0.015) & 0.25(0.19) & 28.6/21 & 12.4 \\
    \hline
  \end{tabular}
  \caption{
    Results of the first order fit ($G_E^s\propto Q^2, G_M^s=$constant).
    Rows are ordered byt the $Q^2$ of the data. Each row gives
    the results of the fit that includes all data
    in Table~\ref{tab_world_data2} up to, and including, the given $Q^2$.
    Row ``$<$0.11~(GeV/c)$^2$'' represents the fit with low $Q^2$ data
    included up to PVA4-H-b. Columns ``$\chi_{min}^2/\nu$''
    and ``Prob'' are the reduced $\chi^2$ and the
    probability ($\equiv \int_{\chi^2_{min}}^{\infty}P_{\chi^2} d\chi^2$)
    of the fit.
  }
  \label{tab_fit_class1}
\end{table}
\endgroup
Compared to the results in Sec.~\ref{sec_comb_0.1}, extending the
fit to higher $Q^2$ gives consistent values for $G_E^s$ and $G_M^s$.
One also notices that for $Q^2$ beyond 0.164~(GeV/c)$^2$, the fit
quality ($\chi_{min}^2/\nu$ and $\chi^2$ probability) deteriorates
significantly, implying that our lowest order model is no longer
able to capture the true $Q^2$ variation in these two quantities.

To better characterize the data at higher $Q^2$, therefore, one
needs to introduce higher order $Q^2$ dependence to $G_E^s$ and/or
$G_M^s$.
For $\sqrt{Q^2}$ sufficiently below the mass of the kaon, chiral
perturbation theory provides a systematic framework for
characterizing the $Q^2$-dependence of nucleon form factors. In this
context the strange magnetic moment, $\mu_s\equiv G_M^s(0)$, arises
at chiral order $p^2$. The sub-leading $Q^2$-dependence of
$G_M^s(Q^2)$ -- sometimes called the strange magnetic radius --  is
nominally ${\cal O}(p^4)$ and is determined by a combination of
chiral loop contributions and a corresponding LEC. Loop effects also
generate an order $p^3$ contribution to the strange magnetic radius
that is free of any new parameters~\cite{Hemmert:1998}.  However,
this contribution is substantially canceled by the ${\cal O}(p^4)$
loop corrections, resulting in a strong dependence on the ${\cal
O}(p^4)$ LEC~\cite{Hammer:2002ei}. In contrast, $G_E^s$ starts off
at chiral order $p^3$, while the sub-leading $Q^2$-dependence (and
corresponding LEC) arises at ${\cal O}(p^5)$.

Based on these considerations, we extend the previous analysis to
include all known, sub-leading $Q^2$-dependence in $G_{M,E}^s$
through ${\cal O}(p^4)$. In practice, doing so amounts to including
 one new constant in our fit associated with the strange magnetic
radius. Since we are interested in the implications for the values
of $G_{M,E}^s$ at $Q^2=0.1$ (GeV/$c$)$^2$, we expand about this
value of $Q^2$ rather than about $Q^2=0$, {\em viz},
\begin{equation}
  \label{eq_param_2ndorder}
  G_E^s(Q^2) = G_E^s\frac{Q^2}{0.1},\quad G_M^s(Q^2) = G_M^s + \mu_s' (Q^2-0.1)\,,
\end{equation}
where $G_E^s$ and $G_M^s$ again are used to represent their values
at $Q^2=$0.1~(GeV/c)$^2$. This will be referred to as the second
order fit. As in the previous section, each row in
Table~\ref{tab_world_data2} now leads to a new constraint in the
form of Eqn.~\ref{eq_least_square_recipe}, with $G_E^s$ and $G_M^s$
parametrized by Eqn.~\ref{eq_param_2ndorder}. Again using
Eqns.~\ref{eq_least_square_recipe} and \ref{eq_chi2_corre}, we
constructed a proper $\chi^2$ function and performed the least
square fit for $(G_E^s,G_M^s,\mu_s')$ by including higher $Q^2$
data. 

The results of this procedure are summarized in
Table~\ref{tab_fit_class3}.
\begingroup
\squeezetable
\begin{table}[!ht]
  \renewcommand{\arraystretch}{1.4}
  \addtolength{\tabcolsep}{-0.1mm}
  \centering
  \begin{tabular}{cccccc} \hline
    $Q^2$
    & \multirow{2}{*}{$G_E^s$} & \multirow{2}{*}{$G_M^s$}
    & $\mu_s'$
    & \multirow{2}{*}{$\chi_{min}^2/\nu$}
    & Prob \\
    (GeV/c)$^2$ & & & (GeV/c)$^{-2}$ &  & (\%) \\\hline
    $<$0.11 & -0.002(0.017) & 0.37(0.25) & -23.7(27.5) & 3.9/4 & 41.3\\\hline
    0.122  & -0.006(0.017) & 0.30(0.24) & -5.9(19.9) & 4.8/5 & 43.6\\
    0.128 & -0.011(0.017) & 0.20(0.23) & 15.7(13.9) & 7.1/6 & 30.7\\
    0.136 & -0.011(0.016) & 0.21(0.23) & 15.7(9.5) & 7.1/7 & 41.4\\
    0.144 & -0.007(0.016) & 0.30(0.22) & 0.1(6.9) & 13.0/8 & 11.0\\
    0.153 & -0.007(0.016) & 0.30(0.22) & -0.1(5.3) & 13.0/9 & 16.1\\
    0.164 & -0.006(0.016) & 0.32(0.22) & -2.3(4.0) & 13.5/10 & 19.9\\
    0.177 & -0.005(0.016) & 0.35(0.22) & -5.8(2.7) & 15.0/11 & 18.4\\
    0.192 & -0.006(0.016) & 0.33(0.22) & -4.2(2.0) & 15.7/12 & 20.6\\
    0.210 & -0.006(0.016) & 0.33(0.22) & -3.6(1.4) & 15.9/13 & 25.3\\
    0.230 & -0.012(0.016) & 0.35(0.21) & -1.5(1.0) & 20.1/14 & 12.6\\
    0.232 & -0.013(0.016) & 0.35(0.21) & -1.6(0.9) & 20.2/15 & 16.5\\
    0.262 & -0.012(0.016) & 0.29(0.21) & -0.4(0.7) & 24.4/16 & 8.2\\
    0.299 & -0.012(0.016) & 0.31(0.21) & -0.6(0.6) & 24.8/17 & 10.0\\
    0.344 & -0.012(0.016) & 0.28(0.20) & -0.3(0.5) & 25.7/18 & 10.8\\
    0.410 & -0.014(0.016) & 0.25(0.20) & 0.1(0.3) & 27.0/19 & 10.4\\
    0.477 & -0.014(0.015) & 0.28(0.20) & -0.1(0.2) & 28.4/20 & 10.1\\\hline
    $<$0.11 and 0.477  & -0.008(0.016) & 0.26(0.22) & -0.3(0.3) & 4.7/5 & 45.7\\
    \hline
  \end{tabular}
  \caption{
    Results of the second order fit ($G_E^s\propto Q^2$
    and $G_M^s$ linear in $Q^2$).
    See the caption of Table~\ref{tab_fit_class1} for a description
    of the table content. Row ``$<0.11$'' represents the fit
    using low $Q^2$ data up to PVA4-H-b. The data at $Q^2$=0.230
    and 0.477~(GeV/c)$^2$
    are from PVA4 and HAPPEx-H-99, respectively, and all others are
    from $G^0$. Uncertainties of $G_E^s$, $G_M^s$, and $\mu_s'$
    are ``marginalized'' 1$\sigma$ uncertainties.
    The last row represents the fit by using only the low $Q^2$ ($<$0.11~(GeV/c)$^2$)
    and HAPPEx-99 data.
  }
  \label{tab_fit_class3}
\end{table}
\endgroup
Several observations can be made from the table. First, compared to
the first order fits, the additional free parameter $\mu_s'$ does
not improve the fit quality significantly. Reasonable fits can be
obtained up to $Q^2$ of 0.210~(GeV/c)$^2$, beyond which the
``flexibility'' of our fit model again seems to be inadequate to
describe the data. Second, the best values of $G_E^s$ and $G_M^s$
are very similar to those obtained from the first order fit
(Table~\ref{tab_fit_class1}). Third, the uncertainties of $G_E^s$
and $G_M^s$, as compared to those from the first order fit, are
slightly larger due to the additional parameter $\mu_s'$. Fourth, if
we ignore the fit quality, and simply examine the mean and
uncertainty of $\mu_s'$, it is large and uncertain until the fit
range goes up to 0.262~(GeV/c)$^2$. This is expected, since it is
difficult to determine the slope parameter with insufficient ``lever
arm''. Also, fits beyond 0.262~(GeV/c)$^2$ suggests a gentle
$\mu_s'$. (To illustrate this point, in the last row in
Table~\ref{tab_fit_class3} we show the fit results by using only the
low $Q^2$ ($<$0.11~(GeV/c)$^2$) and HAPPEx-99 data, which yield a
consistent picture as described above.) However, the fit quality for
$Q^2>$~0.210~(GeV/c)$^2$ prevents us from making a strong statement
about $\mu_s'$ here.

For completeness, we also investigated the impact of including a $Q^4$
term in $G_E^s$ (corresponding to chiral order $p^5$). The resulting 
fit quality does not improve substantially with the inclusion of data
with $Q^2$ beyond $\sim$0.2~(GeV/c)$^2$, and both second order 
parameters in the form
factors are poorly constrained. Nevertheless, the resulting
$G_{E,M}^s$ are consistent with those obtained above.

Based on these considerations, we choose to use the first order fit
up to $Q^2=$0.164 (Table~\ref{tab_fit_class1}) as our final results
in this analysis:
\begin{equation}
\label{eq_bestfit_band2}
G_E^s = -0.008 \pm 0.016 \,, \quad G_M^s = 0.29 \pm 0.21\,,
\end{equation}
with a correlation coefficient of $-0.85$ between the two, and
$\chi^2_{min}/\nu=13.8/11$. Again, the uncertainties of the two form
factors are marginalized 1$\sigma$ uncertainties corresponding to
the projections of $\Delta\chi^2=1$ contour. This result is in good
agreement with Eqn.~\ref{eq_bestfit_band}, for which only the lowest
three $Q^2$ bins of $G^0$ data were included. For $G_M^s$ alone, the one-side
confidence integral for a negative $G_M^s$ ($[-\infty,0]$) is
12.3\%.

\section{\label{sec_conclusion} Conclusion}
A combined analysis of the world PV electron scattering data has
been performed to extract the nucleon strange electric and magnetic
form factors $G_E^s$ and $G_M^s$ at $Q^2 =0.1$~(GeV/c)$^2$. Our
treatment is similar to that of Ref.~\cite{Young06}, but utilizes all
available low $Q^2$ data including the recent HAPPEX 
results~\cite{Ani05, Ani05b} and incorporates the uncertainties in the
electromagnetic and axial form factors. We find that the agreement
among different measurements is good and we obtain fits with
acceptable $\chi^2$. Using a simple parametrization of the $Q^2$
variation in $G_E^s$ and $G_M^s$, a satisfactory global fit can be
obtained up to a $Q^2\sim 0.2$~(GeV/c)$^2$. At
$Q^2=0.1$~(GeV/c)$^2$, the confidence integral for $G_M^s <0$ is
12.3\%, so substantially negative values of $G_M^s$ are highly
disfavored by the fit. In addition, our best fit is consistent with
$G_E^s=0$ at $Q^2$=0.1~(GeV/c)$^2$.

\begin{acknowledgments}
The authors would like to thank Professors Elizabeth J.~Beise,
Douglas H.~Beck, and Jens Erler for very helpful discussions. 
This work was supported in part by National Science Foundation Grant No.
PHY-0555674 and U.~S. Department of Energy Contract No.
DE-FG02-05ER41361.
\end{acknowledgments}


\begin{thebibliography}{99}
\bibitem{MRMOneQuark} The one-quark contributions can be further
enhanced by large logarithms associated with running between the
weak and hadronic scales, leading to relative corrections of order
ten percent or more.
\bibitem{GesZero}$G_E^s(0)\equiv 0$, since there is no net strangeness
in nucleon.
\end{thebibliography}
\end{document}